\documentclass[aps,showpacs,prb,twocolumn,floatfix]{revtex4}
\usepackage{amsmath,amssymb,graphicx,bm}
\newcommand{\Res}{\mathop{\rm Res}\nolimits}
\newcommand{\sgn}{\mathop{\rm sgn}\nolimits}

\begin{document}
\title{Spin-symmetric solution of an interacting quantum dot attached to superconducting leads: Andreev states and the $0-\pi$ transition}

\author{V.  Jani\v{s}}\email{janis@fzu.cz}
\affiliation{Institute of Physics, The Czech Academy of Sciences, Na Slovance 2, CZ-18221 Praha 8, Czech Republic}

\author{ V. Pokorn\'y}  
\affiliation{Institute of Physics, The Czech Academy of Sciences, Na Slovance 2, CZ-18221 Praha 8, Czech Republic}
  
  \author{M. \v Zonda}
 \affiliation{Department of Condensed Matter Physics, Faculty of Mathematics and Physics, Charles University in Prague, Ke Karlovu 5, CZ-12116 Praha 2, Czech Republic}
 
\date{\today}


\begin{abstract}
Behavior of Andreev gap states in a quantum dot with Coulomb repulsion symmetrically attached to superconducting leads is studied via the perturbation expansion in the interaction strength.  We find the exact asymptotic form of the spin-symmetric solution for the Andreev states continuously approaching the Fermi level.  We thereby derive a critical interaction at which the Andreev states  at zero temperature merge at the Fermi energy, being the upper bound for the $0-\pi$ transition.  We show that the spin-symmetric solution becomes degenerate beyond this interaction, in the $\pi$ phase, and the Andreev states do  not split unless the degeneracy is lifted. We further demonstrate that the degeneracy of the spin-symmetric state extends also into the $0$ phase in which the  solutions with zero and  non-zero frequencies of the Andreev states may coexist.         
\end{abstract}

\maketitle 

\section{Introduction}
\label{sec:intro}

Nanostructures attached to leads with specific properties display interesting and important quantum effects at low temperatures. Much attention, both from experimentalists\cite{DeFranceschi10} and theorists,\cite{Yeyati11}  has been paid in recent years to a quantum dot with well separated energy levels attached to BCS superconductors. In particular, behavior of the supercurrent  (Josephson current) that can flow through the impurity without any external voltage bias between the two superconducting leads in equilibrium was in the center of interest.\cite{Jarillo06,Jorgensen06,Jorgensen09} The Josephson current is in the non-interacting or weakly interacting dot proportional to the sine of the phase difference between the left and right superconductor. The supercurrent undergoes a transition signaled  by reversal of its sign when it goes through the impurity with a tangible Coulomb repulsion.\cite{vanDam06,Cleuziou06,Jorgensen07,Eichler09,Maurand12,Delagrange15,Delagrange:2016aa} There is a general consensus that the sign reversal of the Josephson current has the origin in a transition from a spin-singlet state ($0$-phase), where the supercurrent is positive, to a spin-doublet state ($\pi$-phase) where the supercurrent is much smaller in value and negative.\cite{Glazman89,Rozhkov99}  Weak-coupling mean-field static solutions relate the $0-\pi$ transition in the Josephson current to a first-order transition from a spin-symmetric to a spin-polarized state.\cite{Vecino03, Rodero12} This Hartree-Fock solution does not contain dynamical fluctuations and its conclusion about the existence of a magnetic order in the $\pi$-phase is not reliable. More advanced, mostly numerical approaches such as numerical renormalization group (NRG) and\cite{Choi04,Oguri04,Pillet13} Monte-Carlo simulations,\cite{Siano04,Luitz10} further (semi)analytic methods based on an expansion around the atomic limit\cite{Glazman89,Novotny05,Meng09},  slave-particles\cite{Clerk00,Sellier05} and functional renormalization group (fRG)\cite{Karrasch08} produce a continuous $0-\pi$ transition at which no spin-symmetry breaking seems apparent. 

Since there is no tangible evidence that the spin symmetry is broken at the $0-\pi$ transition, the spin-symmetric solution should provide the proper framework within which one can describe this transition. The standard many-body perturbation theory is the best way to do it in a controllable way.  We have to set first a criterion for this transition. The  $0-\pi$ transition is an impurity quantum phase transition (QPT) related to the crossing of  the lowest eigenstates of the many-body Hamiltonian.  A spin-singlet ground state with positive supercurrent ($0$-phase) switches to a spin-doublet state with negative supercurrent ($\pi$-phase). \cite{Matsuura77} This criterion is also used in the numerical simulations. Only single-particle excitations of a non-degenerate many-body (ground) state  are, however, accessible within the standard perturbation theory. It was argued\cite{Meng09} and also experimentally observed\cite{Pillet10,Pillet13}  that  the $0-\pi$ transition is associated with continuous vanishing of the energies of the Andreev bound states  (ABS) at the Fermi energy.  

The knowledge of spectral properties of the one-electron propagator on the impurity is needed to determine the energies of the Andreev bound states. Most theoretical methods  are formulated in the Matsubara formalism within which the spectral properties are not directly accessible. A few approaches have tried to address the impurity Green function with real frequencies  and to understand the $0-\pi$ transition from changes of its spectral properties.\cite{Bauer07,Hecht08,Lim08,Meng09,Luitz10,Rodero12,Zonda15}   Due to the proximity effect the gap of the superconducting leads is imposed on the impurity spectrum and the original atomic levels for the electron and the hole of the dot transform to two Andreev (bound and anti-bound) states in the singlet, spin-symmetric phase. The Andreev states are always symmetrically placed around the Fermi energy (center of the gap). At the transition point the existing two ABS from the $0$-phase reach the Fermi energy and are expected to interchange their positions in the $\pi$-phase.\cite{Pillet10,Pillet13} Our recent analysis of the quantum dot in the weak coupling demonstrated that already the perturbation expansion to the second order gives  an unprecedented accuracy in determining the $0-\pi$ transition when compared to the Numerical Renormalization Group.\cite{Zonda15,Zonda:2016aa} This weak-coupling approach is, however,  inconclusive about the behavior of the Andreev states in the strong-coupling regime. 

An ensuing question is whether there always is a critical interaction strength in the spin-symmetric theory at which the energies of ABS continuously vanish, the bound and anti-bound states merge and eventually cross  in the $\pi$-phase. There are estimates in the literature for the $0-\pi$ transition and crossing of the Andreev states by approximate equaling the Kondo temperature $T_{K}$ of the single-impurity Anderson model (SIAM) to the width of the superconducting gap $\Delta$, $T_{K}\approx \Delta$.\cite{Vecino03,Siano04,Choi04,Bauer07,Lim08,Yeyati11} Positioning of the transition is only of order of magnitude and the continuous or discontinuous character of the transition remains hidden with this inaccurate definition. The Kondo temperature is an external parameter extracted from the asymptotic Bethe-Ansatz gapless solution and is not a proper quantity of the model with a gap.  In the weak coupling the Kondo temperature is usually approximated (replaced) by $T_{K}\to \Gamma_{0}/(1 - \partial \Sigma/\partial \omega)$, where the derivative of the self-energy $\Sigma$ is taken at the Fermi energy and $\Gamma_{0}$ is the hybridization strength of the impurity to the leads.\cite{Vecino03,Bauer07}   

The proof of the existence of a critical interaction for the continuous $0-\pi$ transition in the many-body perturbation theory is still missing.  Vanishing of the energies of ABS beyond the weak-coupling regime in the spin-symmetric solution has not yet been demonstrated. It is the aim of this paper to fill this gap and to give a \textit{non-perturbative proof} of the existence of a critical interaction for continuous vanishing of the energies of ABS at the edge of the $0$-phase in the zero-temperature spin-symmetric solution. 

We introduce the model and the Nambu formalism in Sec.~\ref{sec:model}. We build the standard diagrammatic perturbation expansion in the Coulomb repulsion with Nambu spinors in Sec.~\ref{sec:PE}. All the terms of the perturbation expansion in Matsubara frequencies are eventually analytically continued to real frequencies obtaining thereby their spectral representation in Sec.~\ref{sec:SR-AC}. Analytic continuation allows us to separate the singular contributions from the isolated gap states from the regular ones due to the continuous band states. The gap states carry the positive supercurrent while the band states are responsible for the Kondo screening in the strong coupling. We analyze the asymptotic behavior of the full non-perturbative solution with the energies of ABS approaching the Fermi level and determine the critical interaction at which ABS meet,  Sec.~\ref{sec:SR-EABS}. The behavior of ABS in the weak coupling is numerically resolved  in Sec.~\ref{sec:Multiple-EH}. The findings of the present paper are summarized in Sec.~\ref{sec:conclusions}.

\section{Model Hamiltonian and the Nambu formalism}
\label{sec:model}

A single impurity is used to simulate the nanowire with separated energy levels connecting superconducting leads in the experimental setup.\cite{Choi04,Siano04,Luitz12,Pillet13} The Hamiltonian of the system consisting of a single impurity attached to left-right BCS superconductors is
\begin{equation}
\mathcal{H}=\mathcal{H}_{dot}+\sum_{s=R,L}(\mathcal{H}^s_{lead}+\mathcal{H}^s_c)\ ,
\end{equation}
where the impurity Hamiltonian is a single-level atom with the level energy $\pm\epsilon$ for single electron (hole) and Coulomb repulsion $U$
\begin{equation}
\mathcal{H}_{dot}=\epsilon\sum_\sigma d_\sigma^\dagger d_\sigma^{\phantom{\dag}}
+Ud_\uparrow^\dag d_\uparrow^{\phantom{\dag}} d_\downarrow^\dag d_\downarrow^{\phantom{\dag}} \ .
\end{equation}
The Hamiltonian of the leads is 
\begin{multline}
\mathcal{H}^s_{lead}=\sum_{\mathbf{k}\sigma}
\varepsilon(\mathbf{k})c_{s\mathbf{k}\sigma}^\dag c_{s\mathbf{k}\sigma}^{\phantom{\dag}} \\
-\Delta_{s}\sum_\mathbf{k}(e^{i\Phi_s}
c_{s\mathbf{k}\uparrow}^\dag c_{s\mathbf{-k}\downarrow}^\dag+\textrm{H.c.})\end{multline}
with $s = L,R$ denoting left, right lead. Finally, the hybridization term for the contacts reads
\begin{equation}
\mathcal{H}^s_c=-t_s\sum_{\mathbf{k}\sigma}
(c_{s\mathbf{k}\sigma}^\dag d_\sigma^{\phantom{\dag}}+\textrm{H.c.}) \ .
\end{equation}

We introduce the Nambu spinor representation to describe Cooper pairs and anomalous functions breaking charge conservation. We denote spinors on the impurity site
\begin{equation}
\widehat{\phi}^{\phantom{\dagger}}_{\sigma} = 
\begin{pmatrix}
d^{\phantom{\dagger}}_{\sigma} \\ d^{\dagger}_{-\sigma}
\end{pmatrix} \quad , \quad  \widehat{\phi}^{\dagger}_{\sigma} = \begin{pmatrix}d^{\dagger}_{\sigma} & d^{\phantom{\dagger}}_{-\sigma}\end{pmatrix}\ .
\end{equation}

The individual degrees of freedom of the leads are unimportant for the studied problem and can be integrated out. We are left with only the active variables and functions on the impurity. The fundamental object in the perturbation theory  is the one-electron impurity Green function measuring (imaginary) time fluctuations that in the Nambu formalism is a $2\times 2$ matrix  
\begin{multline}
\widehat{G}_{\sigma}(\tau-\tau') \\=-
\begin{pmatrix}
\langle \mathbb T\left[d_\sigma(\tau) d_\sigma^\dag(\tau')\right]\rangle\ , & \langle \mathbb T\left[ d_{-\sigma}^\dag(\tau) d_\sigma^\dag(\tau')\right]\rangle \\[0.3em]
\langle \mathbb T\left[d_\sigma(\tau) d_{-\sigma}(\tau')\right]\rangle \ , & \langle \mathbb T\left[d_{-\sigma}^\dag(\tau) d_{-\sigma}(\tau')\right]\rangle
\end{pmatrix} \\ 
=
\begin{pmatrix}
G_{\sigma}(\tau - \tau')\ ,		& 	\mathcal{G}^{*}_{\sigma}(\tau - \tau') 	\\
\mathcal{G}_{-\sigma}(\tau - \tau') \ ,	& 	G_{-\sigma}^*(\tau - \tau')
\end{pmatrix} 
\end{multline}
correlating appearance of electrons and holes with a specific spin on the impurity.  
We introduced particle $G_{\sigma}, \mathcal{G}_{\sigma}$ and hole $G^{*}_{\sigma}, \mathcal{G}^{*}_{\sigma}$ Green functions for individual spins. They are connected by symmetry relations $G^{*}_{\sigma}(\tau) = - G_{\sigma}(-\tau) = - G_{\sigma}(\tau)^{\dagger}$, $\mathcal{G}_{\sigma}(\tau) = - \mathcal{G}_{-\sigma}(-\tau)$, $\mathcal{G}^{*}_{\sigma}(\tau) = - \mathcal{G}^{*}_{-\sigma}(-\tau)$, and $\mathcal{G}^{*}_{\sigma}(\tau) = \mathcal{G}_{\sigma}(-\tau)^{\dagger}$, since the normal Green function is odd while the anomalous is an even function of the imaginary time.   

The problem can exactly be solved for the impurity without the onsite interaction, $U=0$. Due to energy conservation it is convenient to use the Fourier transform from (imaginary) time to frequency (energy) where the Green function can analytically be continued to complex values.  The matrix of the inverse Green function for a complex energy $z$ with identical left and fright superconductors reads
\begin{equation}
\widehat{G}_0^{-1}(z)= 
\begin{pmatrix}
z[1+ s(z)]-\epsilon\ , & \Delta\cos(\Phi/2) s(z) \\[0.3em]
\Delta\cos(\Phi/2) s(z)\ , & z[1+ s(z)]+\epsilon
\end{pmatrix},
\end{equation}
where 
\begin{equation}
s(z)=\frac{i\Gamma_0}{\zeta}\mathrm{sgn}(\Im z).
\end{equation}
is the ``hybridization self-energy'', that is, a dynamical renormalization of the impurity energy level due to the hybridization to the superconducting leads. We approximated the Green function in the leads by its value  at the Fermi energy and introduced an effective  hybridization strength  $\Gamma_{0}= 2\pi t^{2}\rho_{0}$. We denoted $\Phi = \Phi_{L} - \Phi_{R}$ the difference between the phases of the left and right superconducting leads and $\rho_{0}$ the density of states of the lead electrons at the Fermi energy.   To represent explicitly the hybridization self-energy we introduced a new complex number $\zeta=\xi+i\eta$ derived from the complex energy $z = x + iy$ by a quadratic equation $\zeta^2=z^2-\Delta^2$. Thereby the following convention for the complex square root has been used 
\begin{equation}
\xi\eta=xy, \quad \mathrm{sgn}(\xi) = \mathrm{sgn} (x), \quad \mathrm{sgn}(\eta) = \mathrm{sign} (y)\ .
\end{equation}

\section{Perturbation expansion in the interaction strength}
\label{sec:PE}

\subsection{One-particle functions}
\label{sec:PE-1PGF}

The full inclusion of the Coulomb repulsion on the impurity cannot be exactly (analytically) performed and we hence must resort to approximations. A systematic way to assess the impact of the Coulomb repulsion on equilibrium properties is a renormalized perturbation expansion. The best way to control the individual contributions from the perturbation expansion is to represent them diagrammatically.  

We start with the Nambu spinor of the impurity propagator that we represent by solid lines  decorated by arrows       
\begin{multline}
\begin{pmatrix}
G_{\sigma}(\tau - \tau')\ ,		& 	\mathcal{G}_{\sigma}^*(\tau - \tau') 	\\
\mathcal{G}_{-\sigma}(\tau - \tau') \ ,	& 	G_{-\sigma}^*(\tau - \tau')
\end{pmatrix}\\ =
\begin{pmatrix}
  \quad\includegraphics[width=12mm]{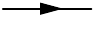}\quad & \quad\includegraphics[width=12mm]{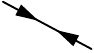}\quad \\
 \quad\includegraphics[width=12mm]{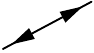}\quad & \quad\includegraphics[width=12mm]{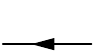}\quad
\end{pmatrix}
\end{multline}
We keep time (charge) propagation (from left to right) in the diagrammatic representation and assign spin to the propagators. Whereby the upper/lower line corresponds to spin up/down.  We can construct standard Feynman many-body diagrams for processes induced by the Coulomb interaction of the electrons on the impurity between two superconducting leads. The Coulomb interaction will be represented via a wavy line. Since the interaction is static, the interaction wavy line is always vertical. Before we start to analyze the diagrammatic contributions from the perturbation expansion we resume basic exact relations.  

The impact of the Coulomb repulsion on the Green function is included in a matrix self-energy $\hat{\Sigma}(z)$ so that the full inverse propagator in the spin-symmetric situation reads 
$\widehat{G}^{-1}(z) =\widehat{G}_0^{-1}(z)-\widehat{\Sigma}(z)$. Its explicit component representation is
\begin{multline}
\widehat{G}^{-1}(z) = \\
\begin{pmatrix}
z[1+s(z)]-\epsilon- \Sigma(z) \ , & \Delta_{\Phi}\left[s(z)- \mathcal{S}(z)\right] \\[0.3em]
\Delta_{\Phi}\left[s(z)-\mathcal{S}^*(z)\right] \ , & z[1+s(z)]+\epsilon - \Sigma^*(z)
\end{pmatrix} \ . 
\end{multline}
We denoted $\Sigma$ and $\mathcal{S}$ the normal and anomalous parts of
the interaction-induced electron self-energy, while $\Sigma^{*}$ and $\mathcal{S}^{*}$ are the self-energies of the hole. The electron-hole symmetry relations
for the unperturbed Green function read in the (complex) energy representation
\begin{equation}\label{eq:EH-G}
G^*(z)=-G(-z)\quad\mathrm{and}\quad \mathcal{G}^*(z)=\mathcal{G}(- z)\ .
\end{equation}
Consequently,  the same relations hold for the self-energies
\begin{equation}\label{eq:EH-Sigma}
\Sigma^*(z)=-\Sigma(- z)\quad\mathrm{and}\quad \mathcal{S}^*(z)=\mathcal{S}(-z) \ .
\end{equation}

If  we denote $D(z)=\det[\hat{G}^{-1}(z)]$, the determinant of the inverse Green function, we then obtain  with the electron-hole symmetry
\begin{multline}\label{eq:D-Sigma}
D(z)=z\left[1 + s(z)\right]\left[z\left(1 +  s(z)\right) - \Sigma(z)  + \Sigma(-z)\right]
\\ 
 - \left[\epsilon + \Sigma(z)\right]\left[\epsilon + \Sigma(-z)\right] -\Delta_{\Phi}^{2}\left[s(z) - \mathcal{S}(z)\right]
\\
\times \left[s(z) - \mathcal{S}(- z)\right]
\end{multline}
and the full one-particle Green function can be represented with the above notation as
\begin{multline}
\widehat{G}(z)=\frac{1}{D(z)}\\ \times
\begin{pmatrix}
z[1+s(z)]+\epsilon + \Sigma(- z)\ , & - \Delta_{\Phi}\left[s(z)-\mathcal{S}(z)\right] \\[0.3em]
-\Delta_{\Phi}\left[s(z)-\mathcal{S}(-z)\right]\ , & z[1+s(z)]-\epsilon- \Sigma(z)
\end{pmatrix}\ .
\end{multline}

The unperturbed ($U=0$) impurity Green function is
\begin{equation}\label{eq:D0}
\widehat{G}^{(0)}(z)=\frac{1}{D_{0}(z)}
\begin{pmatrix}
z[1+s(z)]+\epsilon\ , & -\Delta_{\Phi}s(z) \\[0.3em]
-\Delta_{\Phi}s(z)\ , & z[1+s(z)] - \epsilon
\end{pmatrix}.
\end{equation}
where we denoted $\Delta_{\Phi}=\Delta\cos(\Phi/2) $ and introduced
$$
D_{0}(z) = z^{2}(1+s(z))^{2} - \epsilon^{2} - \Delta_{\Phi}^{2}s(z)^{2} 
$$
the determinant of the matrix of the inverse unperturbed impurity Green function.

\subsection{Two-particle vertex functions}
\label{sec:PE-2PGF}

If we want to go beyond simple weak-coupling approximations such as second order, we have to deal directly with two-particle vertex functions to control the approximations.\cite{Janis14}
We introduce a similar notation for the two-particle vertex to organize the two-particle diagrams in the perturbation expansion. We use the electron-hole notation, which means that the fundamental (normal) vertex contains one electron and one hole. We must assign three independent dynamical variables, Matsubara frequencies in this case, to four end points. We do it for the normal vertex  in the following way
\begin{subequations}\label{eq:xlabel}
\begin{equation}
\label{eq:xlabeln}
\includegraphics[width=8cm]{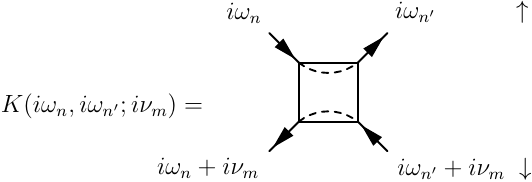}\ .
\end{equation}
The anomalous vertex has a different assignment of frequencies, since the corners of the vertex are connected via diagonals, anomalous Green functions.
\begin{equation}
\label{eq:xlabela}
\includegraphics[width=8cm]{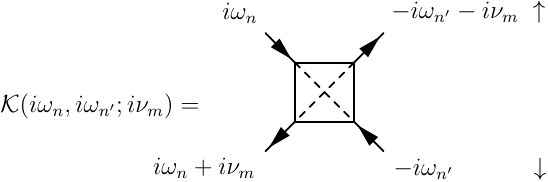}\ .
\end{equation}
\end{subequations}
The convention of attaching the dynamical variables to two-particle Feynman diagrams is as follows. The particles propagate from left to right, the holes from right to left. The arrow indicates propagation of charge. The negative transfer energy $-\nu_{m}$ is the total incoming energy of the particle pair.   

The two-particle vertices $K$ and $\mathcal{K}$ are objects of the perturbation (diagrammatic) expansion that we do not need to specify. We assume that we know them. We connect them with (determine)  the corresponding one-particle self-energy via the Schwinger-Dyson equation. It reads for the normal part of the self-energy 
\begin{equation}
\label{eq:SDE-EHN}
\hspace*{-4pt}\includegraphics[width=9cm]{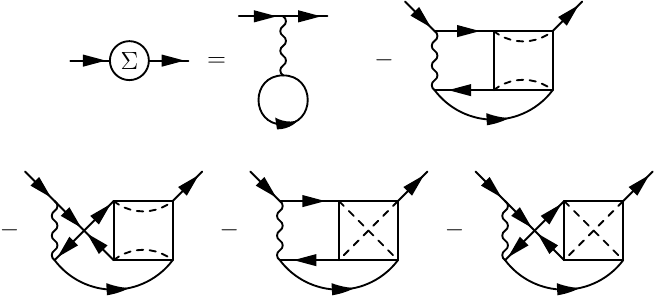} \ ,
\end{equation} 
while the diagrammatic representation for the anomalous self-energy is
\begin{equation}
\label{eq:SDE-EHA}
\hspace*{-4pt}\includegraphics[width=9cm]{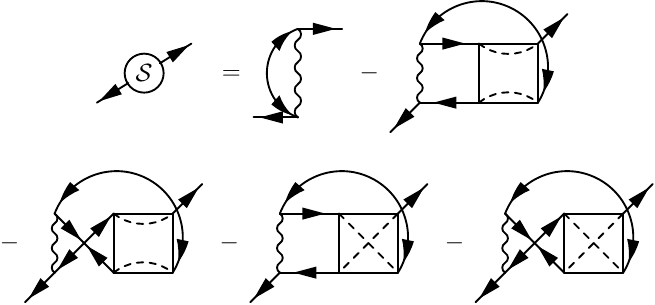}\ .
\end{equation} 
We used dashed lines within the full vertex functions to distinguish normal vertex $K$ (horizontal lines) from the anomalous one $\mathcal{K}$ (diagonal lines). These internal lines indicate the way the corners of the vertices are interconnected by fermionic propagators.

\section{Spectral representation and Andreev bound states}
\label{sec:SR-ABS}

\subsection{Analytic continuation to real frequencies of Matsubara sums}
\label{sec:SR-AC}

General diagrammatic representation with its rules directly gives expressions in Matsubara frequencies. The Matsubara representation on the temperature axis carries no direct information about the gap and the Andreev states. To control the Andreev states and their dependence on the interaction strength we must analytically continue the sums over Matsubara frequencies to spectral integrals over real frequencies. The sums over fermionic and bosonic Matsubara frequencies to be analytically continued generally are
\begin{align}
X(i\nu_m) &=\frac{1}{\beta}\sum_n A(i\omega_n+i\nu_m)B(i\omega_n)\ ,\\
C(i\omega_n) &=\frac{1}{\beta}\sum_m Z(i\nu_m)B(i\omega_n+i\nu_m)\,,
\end{align}
where $\omega_{n}=(2n + 1)\pi T$ and $\nu_{m}=2m\pi T$ are fermionic and bosonic matsubara frequencies, respectively.
The spectral representations of these sums, assuming vanishing of the functions at infinity, are
\begin{multline}
\label{eq:fermionicInt}
X(z)=-\int_{-\infty}^{\infty} \frac{d\omega}{\pi}f(\omega)
\left[\Im A(\omega^+)B(\omega+z) \right. \\ \left. +\Im B(\omega^+)A(\omega-z)\right] 
\end{multline}
and
\begin{multline}
\label{eq:bosonicInt}
C(z)=\mathcal{P}\!\!\int_{-\infty}^{\infty} \frac{d\omega}{\pi}
\left[b(\omega)\Im Z(\omega^+)B(\omega+z) \right. \\ \left. - f(\omega)\Im B(\omega^+)Z(\omega-z)\right]
\end{multline}
where $f(\omega)$ and $b(\omega)$ are Fermi-Dirac and Bose-Einstein distributions.  We abbreviated $\omega^{+} = \omega + i0^{+}$. 

The unperturbed one-electron Green functions have a gap around the Fermi energy  $[-\Delta,\Delta]$ and two poles at $\pm \omega_0$, $ 0\le \omega_0<\Delta$. Since the hybridization self-energy $s(z)$ has a square-root singularity at the gap/band edges, the gap is fixed in the one-electron Green function and does not depend on the interaction strength. The poles and the band edges of the higher-order Green functions do, however, depend on the interaction strength. We hence must be careful when treating the two-particle functions in the spectral representation.

The sum over the fermionic Matsubara frequencies for the one-particle function can then be rewritten in the spectral representation   
\begin{multline}\label{eq:AC-Fermi}
\frac{1}{\beta}\sum_n F(i\omega_n)\rightarrow
-\left[\int_{-\infty}^{-\Delta} + \int_{\Delta}^{\infty}\right] \frac{d\omega}{\pi} f(\omega)\Im F(\omega+i0) \\ +\sum_i f(\omega_i)\Res[F,\omega_i]
\end{multline}
containing an integral over the band states and a sum over isolated poles within the gap.  We first represent the two-particle bubbles via spectral integrals. We resort in further reasoning to zero temperature. The bubble with normal propagators reads
\begin{multline}
\chi(z)=-\int_{-\infty}^{-\Delta}\frac{d\omega}{\pi}
\left[\Im G(\omega^+)G(\omega-z)\right. \\  \left. +\Im G(\omega^+)G(\omega+z)\right]
+\Res[G,-\omega_{0}]G(-\omega_{0}-z)\\+\Res[G,-\omega_{0}]G(-\omega_{0}+z)
\end{multline}
and the anomalous one is 
\begin{multline}
\psi(z)=-\int_{-\infty}^{-\Delta}\frac{d\omega}{\pi}\left[
\Im\mathcal{G}(\omega^+)\mathcal{G}(\omega-z)\right. \\ \left. +\Im\mathcal{G}(\omega^+)\mathcal{G}(\omega+z)\right]
+\Res[\mathcal{G},-\omega_{0}]\mathcal{G}(-\omega_{0}-z)\\  +\Res[\mathcal{G},-\omega_{0}]\mathcal{G}(-\omega_{0}+z) \ .
\end{multline}

The gap of the two-particle bubbles changed, but it is important that it was increased to $[-\Delta-\omega_{0},\Delta+\omega_{0}]$.  The bubbles contain gap states, poles at frequencies $\pm 2\omega_0$. The gap extends to higher values by each convolution of the two-particle propagators. Notice that multiple scatterings in the electron-hole channel contain always a sum of the normal and the anomalous bubble. It follows from the electron-hole symmetry, Eqs.~\eqref{eq:EH-G} and~\eqref{eq:EH-Sigma},  that the poles from the normal bubble are exactly compensated by the poles in the anomalous bubble. Consequently, the total two-particle bubble $\chi(z) + \psi(z)$ in the electron-hole channel is free of gap singularities. 

We now use the normal and anomalous two-particle vertices  in the Schwinger-Dyson equations~\eqref{eq:SDE-EHN} and~\eqref{eq:SDE-EHA} to determine the normal and anomalous self-energy  $\Sigma$, $\mathcal{S}$. We single out for this purpose  the contribution from the static Hartree-Fock approximation to the self-energy and introduce a new two-particle function  
\begin{multline}\label{eq:X-def}
UX(i\omega_{n},i\nu_m) =\frac{1}{\beta}\sum_{n'}\left[ G(i\omega_{n'})G(i\omega_{n'}+i\nu_m) \right. \\ \left. + \mathcal{G}(i\omega_{n'})\mathcal{G}(i\omega_{n'}+i\nu_m)\right]\left[K(i\omega_{n'},i\omega_{n};i\nu_m) \right. \\ \left. + \mathcal{K}(i\omega_{n'},i\omega_{n};i\nu_m)\right] \ .
\end{multline}
constructed from the normal $K$ and anomalous $\mathcal{K}$ full two-particle vertices with three independent frequencies defined in Eqs.~\eqref{eq:xlabel}. We pulled out the bare interaction from the full two-particle vertex so that function $X(i\omega_{n},i\nu_m)$ reduces to a two-particle bubble in the lowest order of the interaction strength. 
  
From now on we denote $\Sigma(\omega)$ and $\mathcal{S}(\omega)$  the normal and anomalous dynamical self-energies, respectively. That is 
\begin{equation}
\begin{split}
\Sigma(i\omega_n)&= -\frac{U^{2}}{\beta}\sum_m G(i\omega_n+i\nu_m) X(i\omega_{n},i\nu_m) \ ,\\
\Delta_{\Phi}\mathcal{S}(i\omega_n)&=- \frac{U^{2}}{\beta}\sum_m \mathcal{G}(i\omega_n+i\nu_m)X(i\omega_{n},i\nu_m)\ .
\end{split}
\end{equation} 
It is more instructive for our analysis to separate the dynamical self-energy from its static, Hartree-Fock part. 

We analytically continue the equations for the self-energy to the spectral representation with real frequencies. Since each convolution increases the gap of the one-electron propagators, we know that the gap of the two-particle function $X$ is not smaller than that of the two-particle bubbles.  When going over to the spectral representation of the sums over Matsubara frequencies we single out the contribution from the gap states. Using Eq.~\eqref{eq:fermionicInt} we obtain a spectral representation for the normal part of the self-energy 
\begin{multline}\label{eq:SDEn}
\Sigma(z) = - U^{2}\Res[G,-\omega_{0}]X(z, -\omega_{0}-z) f(-\omega_{0}) \\ - U^{2}\Res[G,\omega_{0}]X(z, \omega_{0}-z) f(\omega_{0}) + \Sigma_{r}(z) \ ,
\end{multline}
where we denoted the contribution to the self-energy from the band states 
\begin{multline}
\Sigma_{r}(z) = U^{2}\left[\int_{-\infty}^{-\Delta} +\int_{\Delta}^{\infty} \right] \frac{d\omega}{\pi} f(\omega)\Im G(\omega^+)X(z, \omega-z) \\ - U^{2}
\oint_{C}\frac{d \zeta}{2\pi i} b(\zeta)G(z + \zeta)X(z,\zeta) \ .
\end{multline}
Since the analytic structure of function $X(z, \zeta)$ is generally unknown,
the contour $C$ of the second integral is defined implicitly so that to involve the bosonic Matsubara frequencies and avoiding singularities of $G(z + \zeta)X(z,\zeta)$ in variable $\zeta$.  
For the anomalous self-energy we obtain analogously
\begin{multline}\label{eq:SDEa}
\Delta_{\Phi}S (z) 
= - U^{2}\Res[\mathcal{G},-\omega_{0}] X(z, -\omega_{0}-z) f(-\omega_{0}) \\ - U^{2}\Res[\mathcal{G},\omega_{0}]X(z, \omega_{0}-z) f(\omega_{0}) + \Delta_{\Phi}\mathcal{S}_{r}(z)   \ .
\end{multline}
with 
\begin{multline}
\Delta_{\Phi}\mathcal{S}_{r}(z) 
\\
= U^{2}\left[\int_{-\infty}^{-\Delta} +\int_{\Delta}^{\infty} \right] \frac{d\omega}{\pi} f(\omega)\Im \mathcal{G}(\omega^+) X(z, \omega-z) \\ - U^{2}
\oint_{C}\frac{d \zeta}{2\pi i} b(\zeta)\mathcal{G}(z + \zeta)X(z,\zeta) \ .
\end{multline}

We outlined the scheme how individual contributions to the perturbation theory expressed in Matsubara frequencies can be transformed to spectral representations with functions along the axis of real frequencies where we separate the contribution due to the Andreev bound states from those due to the band states. The convolutions generally split into residues of isolated poles from the energy gap and cuts along the continuum of the band states. Such a decomposition is stable in the perturbation expansion, since the gap edges of the one-particle propagator are independent of the interaction strength and the convolutions can only increase the gap in higher-order functions.
 
\subsection{Energies of Andreev states}
\label{sec:SR-EABS}


Positions of the Andreev states are explicitly known in the non-interacting case. One expects crossing of these states when the Coulomb repulsion is strong enough. To demonstrate such a behavior one must determine dependence of the energies of the Andreev states on the interaction strength. We know that the energies of the Andreev states are determined by zeros of the determinant of the inverse of the matrix of the full one-electron Green function. Since ABS lie in the gap, the determinant is real and reads for the general spin-symmetric solution    
\begin{multline}\label{eq:D-Sigma-n}
D(\omega)=\omega\left[1 + s(\omega)\right]\left[\omega\left(1 + s(\omega)\right) - \Sigma(\omega) + \Sigma(-\omega) \right]
\\ 
- \left[\epsilon + Un + \Sigma(\omega)\right]\left[\epsilon + Un + \Sigma(- \omega)\right] \\ -\Delta_{\Phi}^{2}\left[s(\omega) - U\nu - \mathcal{S}(\omega)\right] \left[s(\omega) - U\nu - \mathcal{S}(-\omega)\right]  \ .
\end{multline}

The renormalized energy $\zeta$  along the real axis $z=\omega\pm i0$ is real outside the energy gap $(-\Delta,\Delta)$ and imaginary within it
\begin{equation}
\begin{split}
\zeta&=\sgn(\omega)\sqrt{\omega^2-\Delta^2}\qquad\mathrm{ for }\qquad |\omega|>\Delta,\\
\zeta&=\pm i\sqrt{\Delta^2-\omega^2}\qquad\qquad \!\! \mathrm{ for }\qquad |\omega|<\Delta \ .
\end{split}
\end{equation}
The hybridization self-energy is purely imaginary outside the gap and real within it
\begin{equation}
\begin{split}
s(\omega\pm i0) &=\pm\frac{i\Gamma_0\sgn(\omega)}{\sqrt{\omega^2-\Delta^2}}\qquad\mathrm{ for }\qquad|\omega|>\Delta\ ,
\\
s(\omega\pm i0) &=\phantom{\pm}\frac{\Gamma_0}{\sqrt{\Delta^2-\omega^2}}\qquad\mathrm{ for }\qquad|\omega|<\Delta\ .
\end{split}
\end{equation}

 Zeros of determinant $D(\omega)$  define the frequencies of the  Andreev states. It is easy to find the defining equation for the energy $\omega_{0}$ of the Andreev states for the non-interacting dot. From Eq.~\eqref{eq:D0}   
we directly obtain
\begin{equation}
\omega_{0}(1+ s_{0}) = \pm \sqrt{\epsilon^{2} + \Delta_{\Phi}^{2}s_{0}^{2}} \ .
\end{equation}
We abbreviated $s_{0}=s(\omega_{0})$.


To determine the contributions to the self-energy we need to evaluate the residues of the Green functions in the gap. For the normal and anomalous Green functions the residues at frequency $\omega$ are 
\begin{subequations}\label{eq:HF-residues}
\begin{align}
\Res[G,\omega]&=\frac{\omega\left(1+ s(\omega)\right)+\epsilon+Un + \Sigma(-\omega)}{K(\omega)},\\ 
\Res[\mathcal{G},\omega]&=-\frac{\Delta_\Phi \left[s(\omega) -U\nu - \mathcal{S}(-\omega)\right]}{K(\omega)}\ .
\end{align}\end{subequations}
We denoted the derivative of the determinant $K(\omega)=\partial D(\omega)/\partial \omega$ in the gap. Its explicit representation is 
\begin{multline}
K(\omega)=2\omega \left[1 + \frac{\Gamma_{0}}{\sqrt{\Delta^{2} - \omega^{2}}}\right]\left[1 -\frac{\Sigma'(\omega) + \Sigma'(-\omega)}{2}
\right. \\ \left.  
+ \frac{\Gamma_{0}\Delta^{2}}{\left(\Delta^{2} - \omega^{2}\right)^{3/2}} \right]
- \left[\Sigma(\omega) - \Sigma(-\omega) \right]\left[1 + \frac{\Gamma_{0}\Delta^{2}}{\left(\Delta^{2} - \omega^{2}\right)^{3/2}}\right] 
 \\ 
 - \Delta_{\Phi}^{2}\left[\frac{\omega\Gamma_{0}}{\left(\Delta^{2} - \omega^{2}\right)^{3/2}} -  \mathcal{S}^{\prime}(\omega)\right]\bigg( s(\omega) - U\nu - \mathcal{S}(-\omega)\bigg)
\\
 - \Delta_{\Phi}^{2}\left[\frac{\omega\Gamma_{0}}{\left(\Delta^{2} - \omega^{2}\right)^{3/2}} +  \mathcal{S}^{\prime}(-\omega)\right]\bigg( s(\omega) - U\nu - \mathcal{S}(\omega)\bigg)
 \\
 - \big(\Sigma^{\prime}(\omega) - \Sigma^{\prime}(-\omega)\big) \big( \epsilon + Un \big) 
- \left(\Sigma(\omega)\Sigma(-\omega)\right)^{\prime}\,,
\end{multline}
where prime denotes the frequency derivative. Since  $D(-\omega) = D(\omega)$ is an even function of frequency in the spin-symmetric solution,  its derivative is an odd function, $K(-\omega) = - K(\omega)$. We assume analyticity of the self-energy at the Fermi level $E_{F} =0$. The low-energy asymptotics $\omega \to 0$ of $K(\omega)$   then is $K(\omega) \doteq 2\omega K^{\prime}_{0} $ with
\begin{subequations}\label{eq:K-prime}
\begin{multline}\label{eq:K-primefull}
K^{\prime}_{0} 
= \left[ 1 + \frac{\Gamma_{0}}{\Delta}- \Sigma'(0)\right]^{2}   
- \Sigma^{\prime\prime}(0) \big( \epsilon + U n + \Sigma(0)\big)  
\\
+ \Delta_{\Phi}^{2}S'(0)^{2} -  \Delta_{\Phi}^{2}\left[\frac{\Gamma_{0}}{\Delta^{3}}  
- \mathcal{S}^{\prime\prime}(0) \right] \left[\frac{\Gamma_{0}}{\Delta} - U\nu - \mathcal{S}(0) \right]
\end{multline}  
 being a finite number. Since for zero frequency of the Andreev states, $\omega_{0}=0$ also $\epsilon + U n + \Sigma(0) = 0$ and $\Gamma_{0}/\Delta - U\nu - \mathcal{S}(0) = 0$, we have 
\begin{equation}\label{eq:K-prime0}
K^{\prime}_{0}
= \left[ 1 + \frac{\Gamma_{0}}{\Delta} -  \Sigma'(0)\right]^{2}  
   + \Delta_{\Phi}^{2}S'(0)^{2} >0\,.
\end{equation}  
\end{subequations} 
 
Let  $\omega_{0} \ge 0$ denote the positive frequency of the Andreev state in the spin-symmetric equilibrium state. We resort to zero temperature but do not use any simplifying assumptions there.   We denote $s_{0}^{\pm}= s(\pm\omega_{0}) = s_{0}$, $\mathcal{S}_{0}^{\pm} = \mathcal{S}(\pm\omega_{0})$, $X_{0}^{-} = X(-\omega_{0},0)$, and $X_{- 2}^{+} = X( \omega_{0},- 2\omega_{0}) $ in the explicit calculations.  Moreover, the Green function used in the perturbation expansion is renormalized and hence has the same frequencies of the Andreev states as the full Green function, that is $\pm \omega_{0}$. We single out contributions from the Andreev states  to static parameters $n$ and $\nu$ and  the dynamical self-energy at $\omega_{0}$. We obtain after analytic continuation 
\begin{subequations}\label{eq:par-def}
\begin{align}
n^{\phantom{*}}&= 
 \frac{\omega_{0}( 1 + s_{0}) - \epsilon - U n - \Sigma_{0}^{+}}{K_{0}} +  n_{r}, \\
\nu\  &= \frac{s_{0} - U \nu - \mathcal{S}_{0}^{+}}{K_{0}} + \nu_{r} \ , \\
 \Sigma_{0}^{+} & =  \frac{-\omega_{0}( 1 + s_{0}) + \epsilon + U n + \Sigma_{0}^{+}}{K_{0}}\ U^{2}X_{-2}^{+}  + \Sigma_{0r}^{+}\ , \\
  \Sigma_{0}^{-} & =  \frac{-\omega_{0}( 1 + s_{0}) + \epsilon + U n + \Sigma_{0}^{+}}{K_{0}}\ U^{2} X_{0}^{-} + \Sigma_{0r}^{-} \ , \\
 \mathcal{S}_{0}^{+} & = \frac{U \nu + \mathcal{S}_{0}^{+} - s_{0}}{K_{0}}\  U^{2} X_{-2}^{+}  +  \mathcal{S}_{r0}^{+} \ ,\\
 \mathcal{S}_{0}^{-} & = \frac{U \nu + \mathcal{S}_{0}^{+} - s_{0}}{K_{0}}\ U^{2} X_{0}^{-}  +  \mathcal{S}_{r0}^{-}  
\end{align}
\end{subequations}
where we abbreviated function values $F_{0}= F(\omega_{0})$. We hid  the contributions from the integrals over the band states in unspecified non-singular corrections $n_{r}, \nu_{r}, \Sigma_{r0}^{\pm}$, and $\mathcal{S}_{r0}^{\pm}$. Notice that generally $\omega_{0}(1 + s_{0}) - \epsilon - Un - \Sigma_{0}^{+} \ge 0$ as well as $s_{0}- U\nu - \mathcal{S}_{0}^{+} \ge 0$.  The former expression determines the averaged density of electrons and the latter is proportional to the density of Cooper pairs on the renormalized impurity level $\omega_{0}$. Both must be nonnegative.  It means that the residue cannot cross zero by increasing the interaction strength. They can only reach zero and stay saturated there. That is why the limit $\omega_{0}\to 0$ is of importance. The denominator $K_{0}\to 0$ and the contributions from the gap states become dominant. We may  single them out from the band states. 

We decouple and resolve Eqs.~\eqref{eq:par-def} 
\begin{subequations}\label{eq:param}
\begin{align}\label{eq:param-n}
n &= \frac{\widetilde{\omega}_{0} - \left[\epsilon + U^{2}X_{-2}^{+}n_{r} + \Sigma_{r0}^{+}\right] + K_{0}n_{r}}{U  + K_{0} - U^{2}X_{-2}^{+} }\ , 
\end{align}
\begin{align}
\Sigma_{0}^{+}& = \frac{-\widetilde{\omega}_{0} + \epsilon + U n_{r} +  \Sigma_{r0}^{+}}{U + K_{0} - U^{2} X_{-2}^{+} }\ U^{2}X_{-2}^{+}  + \Sigma_{r0}^{+}\, ,\\
\Sigma_{0}^{-}& = \frac{-\widetilde{\omega}_{0} + \epsilon + U n_{r} +  \Sigma_{r0}^{+}}{U + K_{0} - U^{2} X_{-2}^{+} }\ U^{2}X_{0}^{-}  + \Sigma_{r0}^{-}\ , \\
\nu & =\frac{s_{0}   + \left(K_{0} - U^{2}X_{-2}^{+}\right)\nu_{r} -  \mathcal{S}_{r0}^{+} }{U + K_{0} - U^{2}X_{-2}^{+} } \ , \\
\mathcal{S}_{0}^{+} & =\frac{ U\nu_{r} + \mathcal{S}_{r0}^{+} - s_{0}}{U + K_{0}- U^{2}X_{-2}^{+}  }\ U^{2}X_{-2}^{+}+ \mathcal{S}_{r0}^{+} \, ,
\\
\mathcal{S}_{0}^{-} & =\frac{ U\nu_{r} + \mathcal{S}_{r0}^{+}  - s_{0}}{U + K_{0}- U^{2}X_{-2}^{+}  }\ U^{2}X_{0}^{-}+ \mathcal{S}_{r0}^{-}    \ ,
\end{align}
\end{subequations}
where we denoted $\widetilde{\omega}_{0} = \omega(1 + s_{0})$.

We use the solutions of the equations for $n$ and $\nu$ to derive an explicit equation for the frequencies of the Andreev states. 
It is easy to find
\begin{subequations}\label{eq:sums}
\begin{multline}
\epsilon + Un + \Sigma_{0}^{+}  
\\
\doteq \omega_{0}\left[1 + \frac{\Gamma_{0}}{\Delta} + 2 K'_{0}\frac{\epsilon + U n_{r} + \Sigma_{r0}}{U\left(1 - U X_{00} \right)} \right]\  , 
 \end{multline}
\begin{align}
s_{0} - U\nu - \mathcal{S}_{0}^{+} &\doteq   2\omega_{0} K'_{0}
 \frac{s_{0} - U\nu_{r} -  \mathcal{S}_{r0}^{+}}{U\left(1 - UX_{00}\right) } \, .  
 \end{align}
\end{subequations}
We used the leading-order asymptotics $\omega_{0}\to0$ with $\Sigma_{r0}^{+} - \Sigma_{r0}^{-} = 2\omega_{0}\Sigma_{r}(0)^{\prime}$, $\mathcal{S}_{r0}^{+} - \mathcal{S}_{r0}^{+} = 2\omega_{0}\mathcal{S}_{r}(0)^{\prime} $, and $X_{0}^{-} - X_{-2}^{+}= 2 \omega_{0} \left(\partial_{R} - \partial_{L} \right) X(\omega_{L},\omega_{R})|_{\omega_{L,R}=0} $, where $\partial_{L,R}$ indicates the partial derivative in the left ($\omega_{L}$), right ($\omega_{R}$) variable, respectively. The only assumption made is analyticity of the self-energy $\Sigma(\omega)$  and of the reduced vertex $X(\omega_{L},\omega_{R})$ at the Fermi energy $E_{F} =0$. Determinant $D(\omega_{0})$ is then proportional to $\omega_{0}^{2}$ in the limit $\omega_{0}\to 0$. 
We further need to single out the contribution from the gap states to the derivatives of the normal and anomalous self-energies. Using the representations from Eqs.~\eqref{eq:sums} we easily obtain
\begin{align}
\Sigma'(0) &= \Sigma'_{r0}  +  U \ \frac{\epsilon + U n_{r} +  \Sigma_{r0}}{1 - UX_{00}}\left(\partial_{L} - \partial_{R}\right)X_{00}\, ,\\
\mathcal{S}^{\prime}(0) & =  \mathcal{S}_{r0}^{\prime} 
  - \ U\  \frac{\Gamma_{0}/\Delta - U\nu_{r} -  \mathcal{S}_{r0}}{1  - UX_{00} }\left(\partial_{L} - \partial_{R}\right)X_{00}
\end{align}

 We can now write an equation for the leading-order asymptotics of the determinant $D(x)$ from Eq.~\eqref{eq:D-Sigma-n} in the limit ov vanishing frequencies of ABS, $x=\omega_{0}\to 0$ 
\begin{widetext}
\begin{multline}\label{eq:D0-gen}
\frac{D(\omega_{0})}{4\omega_{0}^{2}} = 
 - \frac{\epsilon + U n_{r} + \Sigma_{r}^{0}}{U(1  - UX_{00})}\ K_{0}^{\prime}\left\{1  + \frac{\Gamma_{0}}{\Delta}  - \Sigma_{r0}^{\prime} + \frac{\epsilon + U n_{r} + \Sigma_{r}^{0}}{U(1  - UX_{00})} \left[K_{0}^{\prime} - U^{2}\left( \partial_{L} - \partial_{R}\right)X_{00} \right]\right\}  
\\
  -  \Delta_{\Phi}^{2} K_{0}^{\prime}\ \frac{\Gamma_{0}/\Delta - U\nu_{r} - \mathcal{S}_{r}^{0}}{U(1  - UX_{00})} \left\{ \frac{\Gamma_{0}/\Delta - U\nu_{r} - \mathcal{S}_{r}^{0}}{U(1  - UX_{00})}\left[K_{0}^{\prime} - U^{2}\left( \partial_{L} - \partial_{R}\right)X_{00}\right]   -  \mathcal{S}_{r}^{0\prime}\right\}  + O(\omega^{2}_{0}) \ .
\end{multline}
\end{widetext}
We denoted the values of the one-electron functions at the Fermi energy with an upper index $0$. Further on, $X_{00}= X(0,0)$.  We see that $\omega_{0}=0$ is a solution for the frequencies of the Andreev states. The expression within the braces is generally nonzero.   The solution $\omega_{0}=0$ exists, however, only if consistency conditions are obeyed, that is $0\le n \le 1$. They explicitly read   
\begin{equation}\label{eq:Omega0-cond}
0 \le -\left[\epsilon + U^{2}X_{00}n_{r} + \Sigma_{r}^{0} \right] \le U\left( 1 - UX_{00}\right) \ .
\end{equation}

The two static parameters in the solution with $\omega_{0}=0$ are 
\begin{subequations}\label{eq:S-omega0}
\begin{align}
n &= - \frac{\epsilon + U^{2}X_{00} n_{r} + \Sigma_{r}^{0}}{U(1  - UX_{00})}\ , \\
\nu &= \frac{\Gamma_{0}/\Delta - U^{2}X_{00}\nu_{r} - \mathcal{S}_{r}^{0}}{U(1 - UX_{00})}\ .
\end{align}
The self-energy for the solution with $\omega_{0}=0$ is 
\begin{align}\label{eq:Sigma_0}
\Sigma(z) &=  - \frac{U^{2}X(z, - z)}{2\left[ 1 + \Gamma_{0}/\Delta + \Sigma_{0}^{\prime}\right]}  + \Sigma_{r}(z) , \\
\mathcal{S}(z) &= \mathcal{S}_{r}(z)\ ,
\end{align}
since the anomalous Green function does not have poles in the gap. We used Eq.~\eqref{eq:K-prime} to determine the residue of the (double) pole in the normal Green function. Equation~\eqref{eq:Sigma_0} induces an equation for $\Sigma_{0}^{\prime}$
\begin{align}
2\left(\Sigma_{0}^{\prime} - \Sigma_{r0}^{\prime}\right)& = \frac{U^{2}\left(\partial_{R} - \partial_{L}\right)X_{00}}{1 + \Gamma/\Delta + \Sigma_{0}^{\prime}} \ .
\end{align} 
\end{subequations}
All equations~\eqref{eq:S-omega0} for static parameters $n,\nu,\Sigma_{0}^{\prime}$ and functions $\Sigma(\omega)$ and $\mathcal{S}(\omega)$ must be solved simultaneously to determine the solution with the Andreev states pinned at the Fermi energy. Such a solution is bounded in the Coulomb repulsion from below, since inequalities in Eq.~\eqref{eq:Omega0-cond} must be obeyed. It may also be bounded from above by a singularity in vertex $X$. Such singularity would indicate a continuous transition to a spin-polarized solution. The solution with $\omega_{0} = 0$ is an isolated point that is very sensitive to magnetic fluctuations. Its stability can be investigated only if external magnetic field is applied and the spin symmetry broken. It is, however, beyond the scope of this paper.   

Notice that the limits $\omega_{0} \to 0$ and $T\to 0$ do not commute. Equations~\eqref{eq:S-omega0} hold for the order $\omega_{0}\to 0$ followed by $T\to 0$. The solution with $\omega_{0}=0$ exists independently of the solution with $\omega_{0}> 0$ if inequalities in Eq.~\eqref{eq:Omega0-cond} are obeyed. The latter solution exists in the weak-coupling regime where the expression on the right-hand side of Eq.~\eqref{eq:D0-gen} is non-zero (negative). With increasing interaction the leading term of the expansion in $\omega^{2}$ of determinant $D(\omega)$  approaches zero. At a critical interaction it vanishes and the Andreev states reach the Fermi energy. Beyond this interaction there is no  solution for (real) nonzero  frequencies of the Andreev states.  The gap states remain frozen at the Fermi energy.  The critical interaction strength at which the Andreev states reach the Fermi energy is determined from the following equation
\begin{widetext}
\begin{multline}\label{eq:Uc-def}
 - U_{c}(1 - U_{c}X_{00}) \left\{ \left[1 + \frac {\Gamma_{0}}\Delta  - \Sigma_{r0}^{\prime}\right]\left(\epsilon + U_{c} n_{r} + \Sigma_{r0} \right) - \Delta_{\Phi}^{2}\left( \frac{\Gamma_{0}}\Delta  - U_{c}\nu_{r} - \mathcal{S}_{r0}\right) \mathcal{S}_{r0}^{\prime} \right\} 
\\ 
= \left\{ \left(\epsilon + U_{c} n_{r} + \Sigma_{r0} \right)^{2} +  \Delta_{\Phi}^{2}\left( \frac{\Gamma_{0}}\Delta  - U_{c}\nu_{r} - \mathcal{S}_{r0}\right)^{2} \right\} \left[K_{0}^{\prime} - U_{c}^{2}\left( \partial_{L} - \partial_{R}\right)X_{00} \right]  \ .
\end{multline}
\end{widetext}
The expression for the critical interaction strength reduces in the Hartree-Fock approximation to
\begin{align}\label{eq:Uc-HFA}
U_{c} &= -\left(1 + \frac{\Gamma_{0}}\Delta \right)\left[ \epsilon + U_{c}n_{r} + \frac{\Delta_{\Phi}^{2}\left( \Gamma_{0}/\Delta - U_{c}\nu_{r}\right)^{2}}{\epsilon + U_{c}n_{r}}\right]\, ,
\end{align} 
which further in the zero-bandwidth limit (generalized atomic limit) simplifies to
\begin{align}\label{eq:Uc-GAL}
U_{c} &= -\left(1 + \frac{\Gamma_{0}}\Delta\right)\left[ \epsilon  + \frac{ \Gamma_{0}^{2}\cos^{2}(\Phi/2)}{\epsilon}\right]\,. 
\end{align}  
We recall that only $\epsilon  < 0$ allows for the existence of a critical interaction. If we further introduce a symmetric notation by using $\xi= \epsilon + U/2$, the critical interaction from Eq.~\eqref{eq:Uc-GAL} reproduces the exact result for the $0-\pi$ transition of the atomic limit,\cite{Meng09} that is for $\Delta\to \infty$,
\begin{align}\label{eq:Uc-atomic}
U_{c} & = 2\sqrt{\xi^{2} + \Gamma_{0}^{2}\cos^{2}(\Phi/2) }\,.
\end{align} 
Notice that the dynamical corrections to the static mean-field are hidden in the derivatives $\Sigma_{r0}^{\prime}$ and $(\partial_{L} - \partial_{R})X_{00}$. Their values depend on the interaction strength and quality of the chosen approximation. They are dominated by the contribution from the band states and contain the information about the Kondo scale. A specific approximation being able to reach the Kondo asymptotics in SIAM should be generalized also to the superconducting quantum dot to decide when the Kondo scale or  temperature dominates the $0-\pi$ transition. Beware that Eq.~\eqref{eq:Uc-def} determines only the upper bound for the $0-\pi$ transition. Only if the spin-symmetric solution remains stable up to the critical interaction $U_{c}$, Eq.~\eqref{eq:Uc-def} determines the real transition to the $\pi$ phase. It seems to be the case in the weak-coupling and atomic limits but it remains unclear in the strong-coupling regime.  

The solution with $\omega_{0}> 0$ develops from the noninteracting state in the $0$-phase. Frequency $\omega_{0}$ is the energy we win when a Cooper pair is created on the dot. This solution at zero temperature  saturates at a critical interaction $U_{c}$ at which the density of Cooper pairs reaches extremum, $s_{0}- U \nu - \mathcal{S}_{0}=0$ and creation and annihilation of Cooper pairs costs no energy. Saturation indicates a degeneracy of the ground state and emergence of a new state. A quantitative description of this  new phase beyond the critical point demands, however, introduction of a perturbation theory for a degenerate ground state. This is beyond the scope of this approach in which we assumed a spin-symmetric and non-degenerate many-body ground state.   

The two solutions, $\omega_{0}> 0$ and $\omega_{0}=0$,  may coexist. The solution with $\omega_{0}> 0$ exists in the weak-coupling region $U < U_{c}$, where the critical interaction $U_{c}$ is determined in Eq.~\eqref{eq:Uc-def}. The solution with $\omega_{0}=0$ exists in the strong-coupling region $U > U_{0}$ where the boundary interaction $U_{0}$ is determined from the right equality in Eq.~\eqref{eq:Omega0-cond}. Both solutions in the coexistence region $U_{0}< U < U_{c}$ are saddle points of the free energy functional. One of them is the true ground state.  The transition between the two solutions at zero temperature is continuous if the weak-coupling state with $\omega_{0}>0$ remains stable up to the critical interaction $U_{c}$. One needs to know the free-energy functional to decide this question and to determine the transition point at non-zero temperatures were the critical interaction $U_{c}(T> 0) =\infty$. 

\section{Andreev states in the weak-coupling regime}
\label{sec:Multiple-EH}
We demonstrate explicitly the universal existence of the critical interaction from the general theory on dynamical approximations in the weak-coupling regime. We use  approximations with a self-energy calculated from second-order perturbation theory or from multiple electron-hole scatterings (RPA). The former is a simplification of the latter. To determine dynamical corrections to the static Hartree-Fock approximation we must evaluate the reduced two-particle vertex $X(i\omega_{n},i\nu_{m})$ defined in Eq.~\eqref{eq:X-def}. The full two-particle vertices $K(i\omega_{n},i\omega_{n'};i\nu_{m})$ and $\mathcal{K}(i\omega_{n},i\omega_{n'};i\nu_{m})$ are determined from Bethe-Salpeter equations. The simplest approximation, the  electron-hole ladder, uses the bare interaction as the the irreducible vertex.  The anomalous irreducible vertex then is zero. The matrix Bethe-Salpeter equation reduces to a couple of algebraic equations for the normal and anomalous vertices $K$ and $\mathcal{K}$.  They are
\begin{subequations}
\begin{multline}
K(i\nu_{m}) = U - U \chi(i\nu_{m}) K (i\nu_{m}) \\ - U \psi(i\nu_{m}) \mathcal{K} (i\nu_{m})
\end{multline}
and
\begin{equation}
\mathcal{K} (i\nu_{m}) = - U \psi(i\nu_{m}) K (i\nu_{m}) - U \chi(i\nu_{m}) \mathcal{K} (i\nu_{m})\ .
\end{equation}
\end{subequations}

The two equations can be decoupled and we obtain explicit solutions
\begin{subequations}\begin{align}
K(i\nu_{m}) &=  \frac{U}{1 + U  \chi(i\nu_{m}) - \displaystyle{\frac{U^{2}\psi(i\nu_{m})^{2}}{1 + U \chi(i\nu_{m})}}}\ , \\
\mathcal{K}(i\nu_{m}) &= - \ \frac{U \psi(i\nu_{m}) }{1 + U \chi(i\nu_{m})} \ K (i\nu_{m})\ .
\end{align}\end{subequations}
The vertex needed for the dynamical self-energy then is
\begin{multline}\label{eq:X-EH}
X(i\omega_{n},i\nu_{m})  =  [\chi(i\nu_{m}) + \psi(i\nu_{m})] \\ \times \frac{1 + U\chi(i\nu_{m}) - U\psi(i\nu_{m})}{[1 + U  \chi(i\nu_{m})]^{2} - U^{2}\psi(i\nu_{m})^{2}} \\ =  \ \frac{\chi(i\nu_{m}) + \psi(i\nu_{m})}{1 + U\left[\chi(i\nu_{m}) + \psi(i\nu_{m})\right]} \ .
\end{multline}
We now use this vertex function to determine the normal and anomalous self-energy, Eqs.~\eqref{eq:SDEn} and~\eqref{eq:SDEa}, that at zero temperature read
\begin{subequations}\label{eq:Sigma-RPA}
\begin{align}
\Sigma(z) &=- U^{2}X(z,-\omega_{0} - z) \Res[G, -\omega_{0}] \nonumber \\ &\qquad + \Sigma_{r}(z)\ , \\
\Delta_{\Phi}\mathcal{S}(z) & = - U^{2}X(z,-\omega_{0} - z) \Res[\mathcal{G}, -\omega_{0}]  \nonumber \\ &\qquad  + \Delta_{\Phi}\mathcal{S}_{r}(z) \ . 
\end{align}
\end{subequations}
They determine together with the equations for the static parameters $n$ and $\nu$ our approximation for $\omega_{0} > 0$. The other solution with $\omega_{0}= 0$ is determined from Eqs.~\eqref{eq:S-omega0}.  

We resolved numerically Eq.~\eqref{eq:X-EH} and Eq.~\eqref{eq:Sigma-RPA} in the weak-coupling regime for interaction strengths with no pole in the two-particle vertex $K(\omega)$. To expedite convergence we used a simplified self-consistency in the above equations.  The one-electron propagators in Eq.~\eqref{eq:X-EH} use a static self-energy of the Hartree-Fock type in a dynamical environment so that the frequencies of the Andreev states are the same as those of the resulting full propagator. The effective dynamical environment is defined by the first iteration of the full dynamical self-consistency. This simplification does not deteriorate the full approximation significantly as demonstrated in Ref.~\onlinecite{Zonda15}. What is of principal importance is the self-consistency in the frequencies of the Andreev states. Otherwise the saturation at the Fermi energy is not reached.       

\begin{figure}
\includegraphics[width=8cm]{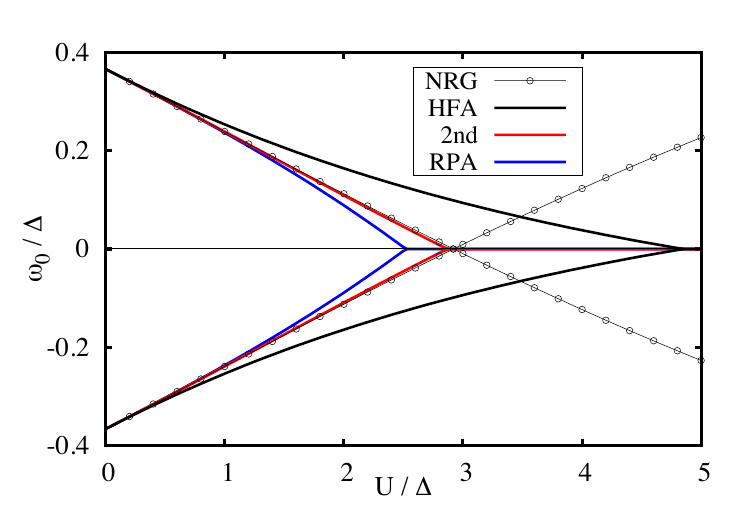}
\caption{(Color online) Frequencies of the Andreev states in the spin-symmetric solution as a function of Coulomb repulsion. Different weak-coupling approximations were compared to demonstrate universality of saturation of the spin-symmetric solution and the NRG data were used as a reference for determining the actual crossing of Andreev states. Input parameters are $\Gamma=\Delta$, $\epsilon= - U/2$ and $\Phi=\pi/2$.  \label{fig:Udep-comp}}
\end{figure}

We calculated the behavior of the Andreev states in three weak-coupling approximations, static Hartre-Fock, dynamical second-order and the ladder of multiple electron-hole scatterings. We also applied Numerical Renormalization Group technique to determine the reference for the behavior of the Andreev states at zero temperature.\cite{NRGLjubljana,Zitko09}  We plotted the frequencies of the Andreev states as a function of the Coulomb repulsion at the charge-symmetric situation in Fig.~\ref{fig:Udep-comp}. We see that all three approximations display the same qualitative behavior derived generally in the preceding section. Only the critical interaction differs in these approximations. We already know that second order approximation gives rather accurate values for the frequencies of the Andreev states,\cite{Zonda15} hence one does not expect significant deviations from them. The difference between the RPA and second order is expected to be compensated by multiple scatterings of electron pairs that were not included in our weak-coupling calculations.  We used only moderate interaction strengths for the zero phase, below the RPA unphysical pole ($U_{c} \approx 3 \sim 4 \Delta$ depending on $\varepsilon$ and $\Phi$) indicating a transition to a magnetic state. 

\begin{figure}
\includegraphics[width=8cm]{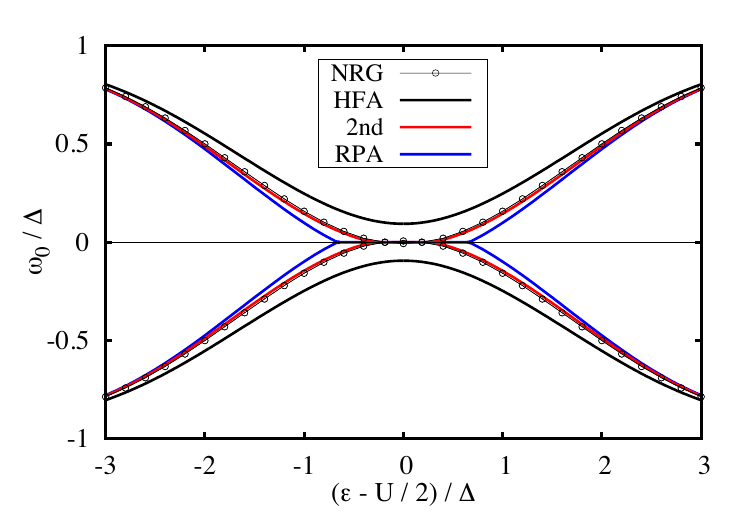}
\caption{(Color online) Frequencies of the Andreev states in the spin-symmetric solution as a function of atomic energy level $\epsilon$. Input parameters are $\Gamma=\Delta$, $U= 3\Delta$ and $\Phi=\pi/2$. The NRG data in the $\pi$-phase are too close to zero to discern the crossing.  \label{fig:Edep-comp}}
\end{figure}

\begin{figure}
\includegraphics[width=8cm]{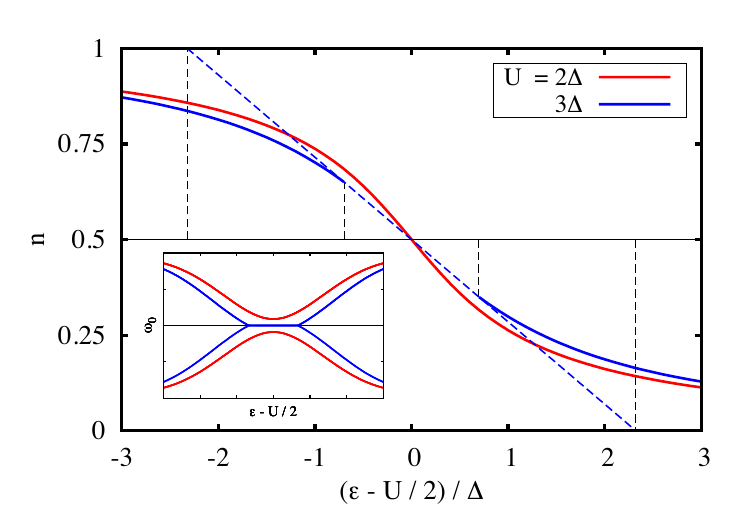}
\caption{(Color online) Electron density as a function of atomic energy level $\epsilon$ for two values of the interaction strength. Limits on solution with $\omega_{0} > 0$ (solid line) and $\omega_{0} = 0$ (dashed line) are indicated. Input parameters are $\Gamma=\Delta$ and $\Phi=\pi/2$. The inset shows the frequencies of the corresponding Andreev states. \label{fig:Endep-comp}}
\end{figure}

\begin{figure}
\includegraphics[width=8cm]{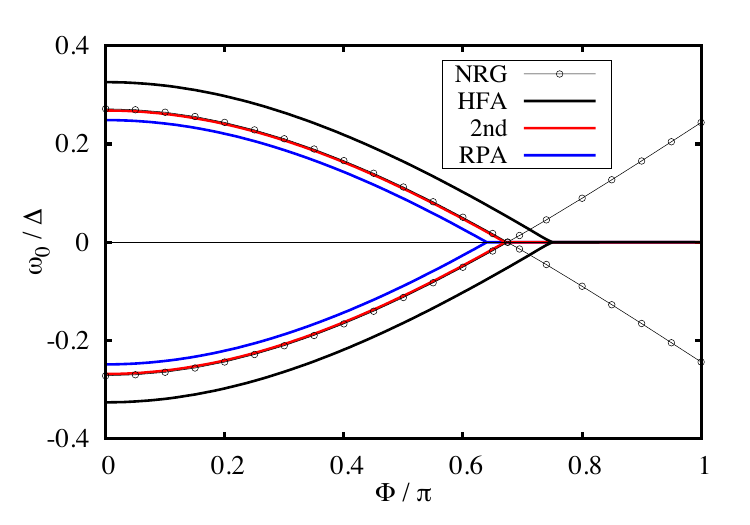}
\caption{(Color online)  Frequencies of the Andreev states in the spin-symmetric solution as a function of phase difference  $\Phi$. Input parameters are $\Gamma=\Delta$, $U= 2\Delta$ and $\epsilon= - \Delta$.  \label{fig:Phidep-comp}}
\end{figure}

The behavior of the Andreev states as a function of the impurity energy level is plotted in Fig.~\ref{fig:Edep-comp}. Here we can see the region where there is no solution with nonzero frequency and only Andreev states with $\omega_{0}=0$ exist. The Hartree-Fock solution does not yet show the saturation at the chosen value of the interaction strength $U = 3\Delta$. The corresponding electron density is plotted in Fig.~\ref{fig:Endep-comp} where we indicated the coexistence region for the solutions with $\omega_{0}> 0$ and $\omega_{0}=0$. The latter solution spans the whole spectrum of electron densities between $n=0$ (right edge) and $n=1$ (left edge). Only the solution of the ladder approximation is plotted.  
The dependence of the frequencies of the Andreev states as a function of the phase difference between the superconducting leads is plotted in Fig.~\ref{fig:Phidep-comp}. The behavior is very similar to that from Fig.~\ref{fig:Udep-comp}.  Increasing the phase difference enhances the effect of the Coulomb repulsion.   

Our numerical calculations illustrated the general and universal conclusion about the behavior of the spin-symmetric solution. The frequencies of the Andreev states approach the Fermi energy with the increasing interaction strength. They merge at the Fermi energy at a critical interaction $U_{c}$ above which only the solution with the Andreev states pinned at the Fermi energy exists. It is important to guarantee in approximate calculations that the frequency of the Andreev states of the Green function used in the equation for the full self-energy, Eq.~\eqref{eq:Sigma-RPA}, are identical with those calculated with the renormalized Green function. If this condition is not fulfilled, fluctuations in the values of the frequencies of the Andreev states in iterations prevent reaching a stable equilibrium state.

\section{Conclusions}
\label{sec:conclusions}

Advanced numerical studies predict crossing of the Andreev bound states at the $0-\pi$  transition without breaking the spin-reflection symmetry.  The aim of our study was to analyze the behavior of the Andreev states in the spin-symmetric solution within the many-body perturbation theory. It is generally known that the Andreev states  tend to approach the Fermi energy with increasing the Coulomb repulsion. We analyzed  the spin-symmetric solution at zero temperature in the asymptotic limit of vanishing energies of ABS with the aim to determine the critical interaction for the $0-\pi$ transition. 
   
We used the diagrammatic expansion in the on-site Coulomb repulsion and assumed a non-degenerate spin-symmetric equilibrium (ground) state. Although formulated in the Matsubara formalism, all the terms of the expansion were analytically continued to real frequencies which allowed us to control the behavior of the Andreev states.   We succeeded in finding the exact asymptotic behavior of the one-electron impurity Green function in the limit of vanishing energy of the Andreev states. We found that the Andreev states at zero-temperature always  reach the Fermi energy at a critical interaction $U_{c}$. Above this critical interaction the Andreev states stay frozen at the Fermi energy. No crossing of the Andreev states can be reached in the spin-symmetric solution with a single ground state. One has to lift the degeneracy of the merged Andreev states in the $\pi$ phase to reproduce the crossing of ABS from the numerical studies. The genesis of the  true $\pi$-phase and the behavior of the Andreev states at and beyond  the transition  can hence be answered in the diagrammatic theory only if  an external magnetic field is applied and stability of the stationary spin-resolved solutions is studied. Breaking the spin-reflection symmetry by the magnetic field does not, however, automatically mean that the solution at zero field becomes magnetic. It is the case only if the linear response of the system to the applied field collapses.  

The solution with the Andreev states pinned at the Fermi energy (zero energy of the Andreev states) can exist independently from the one with separated Andreev states also in the $0$-phase. The existence of the former solution is bounded from below. It can exist only above a certain interaction strength $U_{0} < U_{c}$, and hence the two solutions coexist on an interval $U\in [U_{0},U_{c}]$. The solution with Andreev states  pinned at the Fermi energy is non-perturbative and forms an isolated state within the spin-symmetric phase space.  It is highly sensitive to the external magnetic field that lifts its degeneracy. It is stable only if the linear response to the external magnetic field, breaking the spin-reflection symmetry, is not singular.  The merger of the Andreev states at the Fermi energy is preceded by a discontinuous transition to a magnetic state in the static Hatree-Fock approximation. Stability of solutions with dynamical fluctuations with respect to magnetic fluctuations remains to be investigated. Unless done, the critical interaction from Eq.~\eqref{eq:Uc-def} determines an upper bound for the existence of the spin-symmetric solution, $0$-phase.           

\section*{Acknowledgment} 
We thank T. Novotn\'y   for valuable discussions on the physics of the quantum dot attached to superconducting leads. Research on this problem was supported in part by Grant No. 15-14259S of the Czech Science Foundation (VJ and VP) and by the National Science Centre (Poland) under the contract DEC-2014/13/B/ST3/04451 (M\v Z). 


\end{document}